\documentclass[epj]{svjour}
\usepackage{amsmath}
\usepackage{graphicx}
\begin{document}
\title{On Epsilon Expansions of Four-loop Non-planar Massless Propagator Diagrams}
\author{R.N. Lee\inst{1}
\thanks{e-mail: \texttt{r.n.lee@inp.nsk.su}}%
\and A.V. Smirnov\inst{2,4}
\thanks{e-mail: \texttt{asmirnov80@gmail.com}}
\and V.A. Smirnov\inst{3,4}
\thanks{e-mail: \texttt{smirnov@theory.sinp.msu.ru}}
}                     
%
%
\institute{Budker Institute of Nuclear Physics and
 Novosibirsk State University, 630090, Novosibirsk, Russia \and Scientific Research Computing Center,
Moscow State University, 119992 Moscow, Russia \and Skobeltsyn Institute of Nuclear Physics of Moscow
State University, 119992 Moscow, Russia \and Institut f\"{u}r Theoretische Teilchenphysik, KIT,
76128 Karlsruhe, Germany}
%
%
\abstract{
We evaluate three typical four-loop non-planar massless propagator diagrams
in a Taylor expansion in dimensional regularization parameter $\epsilon=(4-d)/2$
up to transcendentality weight twelve, using a recently developed method
of one of the present coauthors (R.L.).
We observe only multiple zeta values in our results.
} 
\maketitle
%

Analytic results for one-scale multiloop Feynman integrals in a Laurent expansion
in $\epsilon=(4-d)/2$
are expressed as linear combinations of transcendental constants with rational
coefficients. The set of these constants essentially depends on the type of
Feynman integrals. Probably, the simplest type of one-scale Feynman integrals
are massless propagator integrals depending on one external momentum.
Here the world record is set at four loops --- see Ref.~\cite{Baikov:2010hf}
where all the corresponding master integrals were analytically evaluated in
an epsilon expansion up to transcendentality weight seven.

\begin{figure}
  \centering
  \includegraphics[width=7cm]{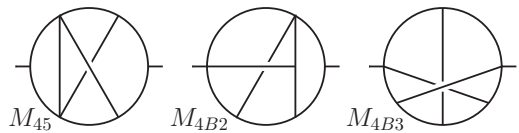}\\
  \caption{Diagrams considered in Ref.~\cite{Brown:2008um} as possible candidates for the presence of sixth
  roots of unity in the $\epsilon$-expansion.}\label{Fig:integrals}
\end{figure}

Practical calculations show that only multiple zeta values (MZV) (see, e.g., \cite{BBV})
appear in results.
Brown proved \cite{Brown:2008um} that convergent scalar massless planar propagator
diagrams with the degree of divergence $\omega\equiv 4 h-2 L=-2$ (where $h$ and $L$ are
numbers of loops and edges, correspondingly) up to five loops contain only MZV
in their epsilon expansions. (In two loops, a proof was earlier presented in Ref.~\cite{BW}.)
He also proved that
for the three non-planar diagrams depicted in Fig.~\ref{Fig:integrals}, every coefficient in a
Taylor expansion in $\epsilon$ is a rational linear combination of
MZV and Goncharov's polylogarithms \cite{Goncharov} with
sixth roots of unity as arguments.

The goal of this brief communication is to study these diagrams experimentally. We present
results in an epsilon expansion up to transcendentality weight twelve. To do this we
apply the DRA method recently suggested by one of the authors (R.L.), Ref. \cite{Lee:2009dh}.
The method is  based on the use of dimensional recurrence relations (DRR) \cite{Tarasov1996} and analytic
properties of Feynman integrals as functions of the parameter of
dimensional regularization, $d$, and was already successfully applied in previous calculations
\cite{Lee:2010cga,Lee:2010ug,Lee:2010wea,Lee:2010hs,Lee:2010ik}.
To apply this method it is essential to perform an integration by parts (IBP)
\cite{IBP} reduction of
integrals that participate in dimensional recurrence relations to master integrals.
To do this, we use the {\tt C++} version of the code {\tt FIRE} \cite{FIRE}.

To study analytic properties of solutions of dimensional recurrence relations,
i.e. to reveal the position and the order of poles in $d$
in a basic stripe, we used a sector decomposition \cite{BH,BognerWeinzierl,FIESTA}
implemented in the code {\tt FIESTA} \cite{FIESTA,FIESTA2}.
To fix remaining constants in the homogenous solutions of dimensional
recurrence relations it was quite sufficient for us to use analytic
results for the four-loop massless propagators master integrals \cite{Baikov:2010hf}
(confirmed numerically by {\tt FIESTA} \cite{Smirnov:2010hd}).
Finally, after obtaining results for master integrals in terms of multiple series
we calculated resulting coefficients at powers of $\epsilon$ numerically with a high
precision and then applied the PSLQ algorithm \cite{PSLQ}.
We also applied the code {\tt HPL} \cite{Maitre:2005uu}, a Mathematica code dealing with harmonic polylogarithms \cite{RV:1999ew}
and MZV.

For all the three diagrams of Fig.~1, we performed evaluation up to transcendentality
weight twelve where the basis of transcendental numbers we used includes 48 constants.
Coefficients in our results turn out to be cumbersome, in particular, such are
the coefficients at $\pi^{12}$ in subsequent formulae. Therefore, to achieve
a successful calculation by PSLQ, we were forced to perform
numerical calculations with a high accuracy. In fact, the accuracy of 800 digits
was enough to obtain results by PSLQ. However, to be on the safe side, we checked
our results by numerical evaluation with the accuracy of 1500 digits.

The first diagram of Fig.~1 is a master diagram. This is nothing but $M_{45}$ in Fig.~2 of
Ref.~\cite{Baikov:2010hf} where all the master integrals for four-loop massless
propagators are shown. Following the method of \cite{Lee:2009dh} we needed, first to
calculate lower master integrals $ M_{01}, M_{11}, M_{13}, M_{14}, M_{21}, M_{27}, M_{35}$.
Eventually, we arrived at the following result which is made homogeneously transcendental
by pulling out an appropriate rational function of $\epsilon$ and normalizing it at
the fourth power of the one-loop integral (i.e. $M_{31}$ according to the notation of \cite{Baikov:2010hf}):
\begin{align}
&\frac{M_{45}(4-2\epsilon)}{\epsilon^4 M_{31}(4-2\epsilon)}
=
\frac{(1-2 \epsilon )^3}{1-6 \epsilon}
\Biggl\{
   36 \zeta _3^2-\biggl(-\frac{6}{5} \pi ^4 \zeta _3+378 \zeta _7\biggr) \epsilon
   \nonumber\\
   &+\biggl(-\frac{427 \pi ^8}{1500}+2844 \zeta _3 \zeta _5+\frac{3024 \zeta _{5,3}}{5}\biggr)\epsilon ^2-\biggl(-\frac{22}{3} \pi ^6 \zeta _3
   \nonumber\\
   &+732 \zeta _3^3+3 \pi ^4 \zeta _5+\frac{42458 \zeta _9}{3}\biggr) \epsilon ^3
   +\biggl(-\frac{60329 \pi ^{10}}{24948}-\frac{183}{5} \pi ^4 \zeta _3^2
   \nonumber\\
   &+62403 \zeta _5^2+149895 \zeta _3 \zeta _7-\frac{58563 \zeta _{8,2}}{2}\biggr)\epsilon ^4
   -\biggl(-\frac{19817}{500} \pi ^8 \zeta _3
   \nonumber\\
   &+\frac{29578 \pi ^6 \zeta _5}{315}+101952 \zeta _3^2 \zeta _5-\frac{50761 \pi ^4 \zeta _7}{50}-325152 \pi ^2 \zeta _9
   \nonumber\\
   &+\frac{73041423 \zeta _{11}}{20}+\frac{175392}{5} \zeta _3 \zeta _{5,3}-\frac{216768}{5} \zeta _{5,3,3}\biggr)\epsilon ^5
   \nonumber\\
   &+\biggl(-\frac{98988919597 \pi ^{12}}{17027010000}-\frac{19242}{35} \pi ^6 \zeta _3^2+\frac{37216 \zeta _3^4}{3}
   \nonumber\\
   &-\frac{10280}{9} \pi ^4 \zeta _3 \zeta _5+\frac{210112}{3} \pi ^2 \zeta _5^2+\frac{600320}{3} \pi ^2 \zeta _3 \zeta _7
   \nonumber\\
   &+2100799 \zeta _5 \zeta _7+\frac{9455470 \zeta _3 \zeta _9}{9}+\frac{160424}{75} \pi ^4 \zeta _{5,3}
   \nonumber\\
   &+60032 \pi ^2 \zeta _{7,3}-\frac{1132952 \zeta _{9,3}}{3}-120064 \zeta _{6,4,1,1}\biggr)\epsilon ^6
   \nonumber\\
   &+O\left(\epsilon ^7\right)
\Biggr\}\,.
\end{align}

Here $\zeta_{m_1,\dots,m_k}$ are MZV given by
\begin{equation}\label{MZVdef}
\zeta(m_1,\dots,m_k)=
\sum\limits_{i_1=1}^\infty\sum\limits_1^{i_1-1}
\dots\sum\limits_1^{i_{k-1}-1}\prod\limits_{j=1}^k\frac{\mbox{sgn}(m_j)^{i_j}}{i_j^{|m_j|}}\,.
\end{equation}

The other two diagrams of Fig.~1 {\em are not} master integrals and, consequently,
they do not appear in Fig.~2 of Ref.~\cite{Baikov:2010hf}. We found their reduction to lower
master integrals by FIRE and then evaluated resulting master integrals by our technique.
Let us stress that the master integrals involved in the IBP reduction of
the second and the third non-planar diagrams of Fig.~\ref{Fig:integrals} are all {\em planar}
diagrams. However, the result of Brown \cite{Brown:2008um} quoted in the beginning of
our paper is not applicable to such planar diagrams.
(Two of them are $M_{35}$ and $M_{36}$ in Fig.~2 of Ref.~\cite{Baikov:2010hf}, with
$\omega=0$, rather than $-2$.) So, {\em a priori},
one could admit the presence of something additional to MZV in their epsilon expansions.

These are our results for them in the same normalization where we again managed to reveal
homogenous transcendentality:
\begin{align}
&\frac{M_{4B2}(4-2\epsilon)}{\epsilon^4 M_{31}(4-2\epsilon)}
=
\frac{(1-2 \epsilon )^3}{1
   -\epsilon }
\Biggl\{
   36 \zeta _3^2+\biggl(\frac{6 \pi ^4 \zeta _3}{5}+\frac{189 \zeta _7}{2}\biggr) \epsilon
   \nonumber\\
   &+\biggl(\frac{1359 \pi ^8}{7000}+144 \zeta _3 \zeta _5-\frac{2916 \zeta _{5,3}}{5}\biggr)\epsilon ^2+\biggl(\frac{4 \pi ^6 \zeta _3}{21}-1392 \zeta _3^3
   \nonumber\\
   &+51 \pi ^4 \zeta _5+\frac{25549 \zeta _9}{6}\biggr) \epsilon ^3
   +\biggl(\frac{146255 \pi ^{10}}{99792}-\frac{348}{5} \pi ^4 \zeta _3^2
   \nonumber\\
   &-\frac{75843 \zeta _5^2}{4}-\frac{206655 \zeta _3 \zeta _7}{4}+\frac{152163 \zeta _{8,2}}{8}\biggr)\epsilon ^4
   \nonumber\\
   &+\biggl(-\frac{313039 \pi ^8 \zeta _3}{31500}+\frac{25982 \pi ^6 \zeta _5}{315}-39372 \zeta _3^2 \zeta _5+\frac{234339 \pi ^4 \zeta _7}{200}
   \nonumber\\
   &+29412 \pi ^2 \zeta _9-\frac{26995129 \zeta _{11}}{160}+\frac{71928}{5} \zeta _3 \zeta _{5,3}+\frac{19608}{5} \zeta _{5,3,3}\biggr)\epsilon ^5
   \nonumber\\
   &+\biggl(\frac{107930288857 \pi ^{12}}{68108040000}-\frac{42968}{315} \pi ^6 \zeta _3^2+\frac{85696 \zeta _3^4}{3}
   \nonumber\\
   &-\frac{23510}{9} \pi ^4 \zeta _3 \zeta _5+\frac{31822}{3} \pi ^2 \zeta _5^2+\frac{90920}{3} \pi ^2 \zeta _3 \zeta _7+165244 \zeta _5 \zeta _7
   \nonumber\\
   &-\frac{6321395 \zeta _3 \zeta _9}{9}+\frac{45614}{75} \pi ^4 \zeta _{5,3}+9092 \pi ^2 \zeta _{7,3}-\frac{803569 \zeta _{9,3}}{6}
   \nonumber\\
   &-18184 \zeta _{6,4,1,1}\biggr)\epsilon ^6+O\left(\epsilon ^7\right)
\Biggr\}\,,
\end{align}
\begin{align}
&\frac{M_{4B3}(4-2\epsilon)}{\epsilon^4 M_{31}(4-2\epsilon)}
=
\frac{(1-2 \epsilon )^3}{1+4 \epsilon}
\Biggl\{
   36 \zeta _3^2+\biggl(\frac{6 \pi ^4 \zeta _3}{5}+567 \zeta _7\biggr) \epsilon
   \nonumber\\
   &+ \biggl(\frac{211 \pi ^8}{1750}+3744 \zeta _3 \zeta _5+\frac{1944 \zeta _{5,3}}{5}\biggr)\epsilon ^2+\biggl(\frac{68 \pi ^6 \zeta _3}{7}+288 \zeta _3^3
   \nonumber\\
   &+30 \pi ^4 \zeta _5+22094 \zeta _9\biggr) \epsilon ^3
   + \biggl(\frac{1255 \pi ^{10}}{4158}+\frac{72}{5} \pi ^4 \zeta _3^2
   \nonumber\\
   &+51318 \zeta _5^2+95490 \zeta _3 \zeta _7-11799 \zeta _{8,2}\biggr)\epsilon ^4
   + \biggl(\frac{3283 \pi ^8 \zeta _3}{125}
   \nonumber\\
   &+\frac{12484 \pi ^6 \zeta _5}{105}-65952 \zeta _3^2 \zeta _5+\frac{17538 \pi ^4 \zeta _7}{25}+65232 \pi ^2 \zeta _9
   \nonumber\\
   &-\frac{360909 \zeta _{11}}{10}-\frac{54432}{5} \zeta _3 \zeta _{5,3}+\frac{43488}{5} \zeta _{5,3,3}\biggr)\epsilon ^5
   \nonumber\\
   &+\biggl(\frac{2972813873 \pi ^{12}}{1064188125}-\frac{15056}{105} \pi ^6 \zeta _3^2-29608 \zeta _3^4-\frac{3640}{3} \pi ^4 \zeta _3 \zeta _5
   \nonumber\\
   &-8176 \pi ^2 \zeta _5^2-23360 \pi ^2 \zeta _3 \zeta _7+1091724 \zeta _5 \zeta _7+\frac{4198640 \zeta _3 \zeta _9}{3}
   \nonumber
\end{align}
\begin{align}
   &-\frac{6752}{25} \pi ^4 \zeta _{5,3}-7008 \pi ^2 \zeta _{7,3}+95936 \zeta _{9,3}+14016 \zeta _{6,4,1,1}\biggr)\epsilon ^6
   \nonumber\\
   &+O\left(\epsilon ^7\right)
\Biggr\}\,.
\end{align}

We see that only MZV are present in our results.
Although Goncharov's polylogarithms at sixth roots of unity were allowed
to appear according to the analysis of Ref.~\cite{Brown:2008um} they have not
appeared. In fact, such transcendental numbers do appear in epsilon expansions
of some classes of {\em massive} Feynman integrals \cite{B99,MK1,MK2,MK3}.

Taking our results into account one could try to prove that
that there are only MZV in massless propagator diagrams.
Unfortunately, the way Brown proceeded ~\cite{Brown:2008um} in the case of
a few concrete
diagrams can hardly be generalized to other situations because his analysis
was based on the possibility to perform recursive integrations in Feynman parameters,
as long as the denominator of the integrand is linear with respect to them, so that
some other approach is needed. On the other hand, attempts to discover new types of
transcendental numbers, in addition to MZV, meet more and more difficulties because we
have to turn to more loops and more terms of $\epsilon$-expansions.
Still this is certainly possible at the current level of our possibilities so that
we are planning to do this in the nearest future.

\vspace{0.2 cm}

{\em Acknowledgments.}

This work was supported by the Russian Foundation for Basic Research through grant
11-02-01196 and by DFG through SFB/TR~9 ``Computational Particle Physics''.
The work of R.L. was also supported through Federal special-purpose program
``Scientific and scientific-pedagogical personnel of innovative Russia''. R.L. gratefully acknowledges
Karlsruhe Institut f\"{u}r Theoretische Teilchenphysik for warm hospitality and
financial support during his visit. We are grateful to David Broadhurst and Mikhail Kalmykov for
instructive discussions.


\begin{thebibliography}{99}

\bibitem{Baikov:2010hf}
  P.~A.~Baikov and K.~G.~Chetyrkin,
  Nucl.\ Phys.\  B {\bf 837} (2010) 186
  [arXiv:1004.1153 [hep-ph]].

\bibitem{BBV}
J.~Blumlein, D.~J.~Broadhurst and J.~A.~M.~Vermaseren,
  Comput.\ Phys.\ Commun.\  {\bf 181} (2010) 582
  [arXiv:0907.2557 [math-ph]].

\bibitem{Brown:2008um}
  F.~Brown,
  Commun.\ Math.\ Phys.\  {\bf 287}, 925 (2009)
  [arXiv:0804.1660 [math.AG]].

\bibitem{BW}
I.~Bierenbaum and S.~Weinzierl,
Eur.\ Phys.\ J.\  C {\bf 32} (2003) 67.

\bibitem{Goncharov}
 A.~B.~Goncharov,
 ``Multiple polylogarithms, cyclotomy and modular complexes,''
 Math.\ Research Letters {\bf 5} (1998) 497;
``Multiple polylogarithms and mixed Tate motives,''
[arXiv:math/0103059].

\bibitem{Lee:2009dh}
  R.~N.~Lee,
Nucl.\ Phys.\  B {\bf 830} (2010) 474
  [arXiv:0911.0252 [hep-ph]].

\bibitem{Tarasov1996}
O.~V.~Tarasov, Phys.~Rev.~D {\bf 54} (1996) 6479.





\bibitem{Lee:2010cga}
  R.~N.~Lee, A.~V.~Smirnov and V.~A.~Smirnov,
  JHEP {\bf 1004} (2010) 020
  [arXiv:1001.2887 [hep-ph]].

\bibitem{Lee:2010ug}
  R.~N.~Lee, A.~V.~Smirnov and V.~A.~Smirnov,
  Nucl.\ Phys.\ Proc.\ Suppl.\  {\bf 205-206} (2010) 308
  [arXiv:1005.0362 [hep-ph]].

\bibitem{Lee:2010wea}
  R.~N.~Lee,
  Nucl.\ Phys.\ Proc.\ Suppl.\  {\bf 205-206} (2010) 135
  [arXiv:1007.2256 [hep-ph]].

\bibitem{Lee:2010hs}
  R.~N.~Lee and I.~S.~Terekhov,
  JHEP {\bf 1101} (2011) 068
  [arXiv:1010.6117 [hep-ph]].

\bibitem{Lee:2010ik}
  R.~N.~Lee and V.~A.~Smirnov,
  JHEP {\bf 1102} (2011) 102
  [arXiv:1010.1334 [hep-ph]].

\bibitem{IBP}
  K.~G.~Chetyrkin and F.~V.~Tkachov,
  Nucl.\ Phys.\  B {\bf 192} (1981) 159.

\bibitem{FIRE}
 A.~V.~Smirnov,
  JHEP {\bf 0810}, 107 (2008)
  [arXiv:0807.3243 [hep-ph]].

\bibitem{BH}
T.~Binoth and G.~Heinrich, Nucl. Phys. B, {\bf 585} (2000) 741;
Nucl. Phys. B, {\bf 680} (2004) 375;
Nucl. Phys. B, {\bf 693} (2004) 134;
G.~Heinrich, Int. J. of Modern Phys. A,  {\bf 23} (2008) 10.
[arXiv:0803.4177];
J.~Carter and G.~Heinrich,
arXiv:1011.5493 [hep-ph].

\bibitem{BognerWeinzierl}
C.~Bogner and S.~Weinzierl,
Comput. Phys. Commun.  {\bf 178} (2008) 596
[arXiv:0709.4092 [hep-ph]];
Nucl. Phys. Proc. Suppl.  {\bf
183} (2008) 256 [arXiv:0806.4307 [hep-ph]].

\bibitem{FIESTA}
  A.~V.~Smirnov and M.~N.~Tentyukov,
  Comput.\ Phys.\ Commun.\  {\bf 180} (2009) 735
  [arXiv:0807.4129 [hep-ph]].

\bibitem{FIESTA2}
 A.~V.~Smirnov, V.~A.~Smirnov and M.~Tentyukov,
  Comput.\ Phys.\ Commun.\  {\bf 182} (2011) 790
  [arXiv:0912.0158 [hep-ph]].

\bibitem{Smirnov:2010hd}
  A.~V.~Smirnov and M.~Tentyukov,
  Nucl.\ Phys.\  B {\bf 837} (2010) 40
  [arXiv:1004.1149 [hep-ph]].


\bibitem{PSLQ}
 H.~R.~P.~Ferguson, D.~H.~Bailey and S.~Arno,
  Math.\ Comput.\  {\bf 68}, (1999) 351,
  NASA--Ames~Technical Report,
  NAS--96--005.

\bibitem{Maitre:2005uu}
  D.~Maitre,
  Comput.\ Phys.\ Commun.\  {\bf 174} (2006) 222
  [arXiv:hep-ph/0507152].
\bibitem{RV:1999ew}
  E.~Remiddi, J.~A.~M.~Vermaseren,
  Int.\ J.\ Mod.\ Phys.\  {\bf A15} (2000) 725
  [hep-ph/9905237].
\bibitem{B99}
D.~J.~Broadhurst,
Eur.\ Phys.\ J.\  C {\bf 8} (1999) 311.

\bibitem{MK1}
J.~Fleischer and M.~Yu.~Kalmykov,
Phys.\ Lett.\  B {\bf 470} (1999) 168.

\bibitem{MK2}
M.~Yu.~Kalmykov,
Nucl.\ Phys.\  B {\bf 718} (2005) 276.

\bibitem{MK3}
  M.~Yu.~Kalmykov and B.~A.~Kniehl,
  Nucl.\ Phys.\ Proc.\ Suppl.\  {\bf 205-206} (2010) 129
  [arXiv:1007.2373 [math-ph]].

\end{thebibliography}
\end{document}